# Diffraction of light by topological defects in liquid crystals

E. Pereira*[1] and F. Moraes

*Departamento de Física, Universidade Federal da Paraíba, Caixa Postal 5008, 58051-970, João Pessoa, Paraíba, Brazil*

**Abstract**

We study light scattering by a hedgehog-like and linear disclination topological defects in a nematic liquid crystal by a metric approach. Light propagating near such defects feels an effective metric equivalent to the spatial part of the global monopole and cosmic string geometries. We obtain the scattering amplitude and the differential and total scattering cross section for the case of the hedgehog defect, in terms of the characteristic parameters of the liquid crystal. Studying the disclination case, a cylindrical partial wave method is developed. As an application of the previous developments, we also examine the temperature influence on the localization of the diffraction patterns.

## 1 Introduction

Recently, a geometrical model for light propagation in nematic liquid crystals has been put forward by one of us and a collaborator [1]. In this model Fermat's principle is equated to the variational principle that determines geodesics in Riemannian geometry [2]. In doing so, the path described by light is interpreted as a geodesic in an effective geometry, allowing not only a different way of thinking on the subject, but presenting the tools of Riemannian geometry as serviceable ones to produce results that otherwise would be very complicated to get [3]. In the present article we make use of the effective geometrical background in order to study diffraction of light by topological defects in nematics. We look at a disclination and a point defect (hedgehog), each of topological charge +1. The choice is due to their azimuthal symmetry which much simplifies the calculations.

*Corresponding author. Email: erms@fisica.ufpb.br

Topological defects in nematics, as in many physical systems, appear naturally in a symmetry-breaking phase transition. The high temperature isotropic phase has complete 3D rotational symmetry which is reduced to 2D rotational symmetry around the director axis of the nematic phase. Quenching the isotropic phase leads to the formation of domains with different orientations in the nematic phase and, consequently, to the creation of defects. Among these, the commonest are the disclinations and point defects named hedgehogs. Something analogous apparently happened at the beginning of the universe: as the universe expanded and cooled down it went through a sequence of symmetry-breaking phase transitions that left behind topological defects like cosmic strings and monopoles. This has been shown [4] to be analogous, from the point of view of light propagation, to disclinations and hedgehogs. As noticed by Bowick *et al.* [5], the physical principle that conducts the creation of these defects in liquid crystals is the same found in the creation of cosmic strings for the universe: the mechanism of Kibble [6].

There has been much effort put on analogous gravitational systems like the one described above in recent years [7, 8]. Because these systems present similar mathematical description, analogous models suggest the possibility of reproducing selected gravitational systems in the condensed matter laboratory, allowing the study of aspects of general relativity not accessible to the current technology. For example, when a fluid flow has a downstream current, any phonon propagating in this surroundings feels a kind of a gravitational pull and black-hole-like equations rule the phonon's path [9]. In this example, the phonon propagation is described by an effective Riemannian geometry, such that null geodesics of this geometry are traced by the phonons. Similarly, the use of an effective metric, as described in [1], will be done in the present article to study new, wave-like, aspects of the propagation of the light.

In this article we study the propagation of light around topological defects in nematics from the point of view of wave interference and diffraction. We examine what happens to a plane wave which collides with a disclination and a hedgehog defect. We calculate the scattering amplitude for both cases and obtain the respective diffraction pattern. With this we complete the analysis on propagation of light in nematics with topological defects initiated in [1] where the geometric optics approach was used. As it will become clear below, the ondulatory approach suggests a way of measuring the ratio between the ordinary and extraordinary refractive indices only by finding the position of the first diffraction peak. Furthermore, the possibility of building diffraction gratings with arrays of line defects in nematics appears as a practical application. Also, since the problem studied here is analogous to the problem of light scattering off a cosmological defect, we offer a means of simulating a cosmological experiment in a condensed matter laboratory. This way, this article also contributes to the body of knowledge that is being accumulated on condesed matter analogues of gravitational systems [7].

This article is divided as follows. In the first part, we present the geometric model and the effective metrics used throughout this work. In the second and the third parts, we apply the partial wave method to the metrics generated, respectively, by hedgehog and disclination defects. In the hedgehog case, we obtain the diffraction pattern, its localization, the scattering amplitude and the differential and total scattering cross section, while in the disclination case, we emphasize the localization of the diffraction pattern. In the fourth part, we analyze the temperature influence on the localization of the diffraction patterns and, in the fifth one, we present our conclusions.

## 2 Geometric Model

Observing the light propagating between two isotropic media, no new effect in its path is noticed, besides those foreseen by Fresnel's equations. However, considering the passage of a light ray from an isotropic medium to an anisotropic one, for example, a liquid crystal in the uniaxial nematic phase (where the uniaxial attribute is represented by the director vector **n**), one observes a split from the original ray into two other rays with perpendicular polarizations among themselves: the ordinary and extraordinary rays. The former's electric field polarization is perpendicular to **n**, as long as the latter's one has a component lying on **n**. To calculate the propagation of a monochromatic wave, through Maxwell's equation in a medium without currents and charges, one notices [11] that the refractive index for the extraordinary ray is given by

$$N_e^2(\beta) = n_o^2 \cos^2 \beta + n_e^2 \sin^2 \beta \tag{1}$$

where β is the angle between the director **n** and the Poynting vector **S** of the extraordinary ray (that is the actual direction of light propagation because **S** indicates the energy flow), and $n_e$ and $n_o$ are, respectively, the refractive indexes of the rod molecules which constitute the liquid crystal.

With this refractive index, it is possible to calculate the path of the extraordinary light applying Fermat's principle. This says that the path traveled by light, among two points *A* and *B*, will be that which minimizes the integral

$$F = \int_A^B N_e dl, \tag{2}$$

using a parameter *l* along the path. From Riemannian geometry, the light trajectory among two points will be that which minimizes the line element *ds* given by

$$ds^2 = \sum_{i,j=1}^{3} g_{ij} dx^i dx^j , \tag{3}$$

where, $\{i, j\} = \{r, \theta, \phi\}$ in spherical coordinates, and $g_{ij}$ are the components of the metric tensor **g**. Thus, it is possible to derive the geometric model by noticing that both Fermat's principle and the calculus of geodesics in Riemannian manifolds are questions of minimizing physical amounts. Bearing this in mind, if we compare $(N_e dl)^2$, from Fermat's principle, with $ds^2$, from a generic Riemannian manifold, we interpret the path traveled by light as a geodesic in an effective metric, that is, we are led to the following identity

$$N_e^2 dl^2 = \sum_{i,j=1}^{3} g_{ij} dx^i dx^j . \tag{4}$$

This way, it is possible to calculate the effective metric felt by the light finding cosβ and sinβ, in (1) and substituting (1) in (4).

In this article, we study the hedgehog point defect with topological charge +1, with the director **n** = $\hat{r}$, and the $\{\kappa = 1, \gamma = 0\}$ wedge disclination [10]. The constants κ and γ are related to the general wedge disclination's director in Cartesian coordinates and plane $z$ = *const*. by **n** = $(\cos(\kappa\phi + \gamma), \sin(\kappa\phi + \gamma), 0)$ [1, 4, 11]. In laboratory, the hedgehog can be confined inside an isotropically transparent spherical vessel or a nematic droplet [13], while the disclination can be confined inside an isotropically transparent cylindrical vessel [14]. The effect of a transparent isotropic medium (the vessel) on the light will be that of a lens and is neglected here [15,16]. Despite of the common knowledge of the 3D-escape of the director for the (k=1,c=0) disclination, we assume a stable disclination due two reasons. First, it is a experimental reality in liquid crystal polymers [11,17]. Second, this was chosen as starting point to study the assymetrical k≠1 defects due to its simplicity and axial symmetry.

The procedure to find effective metrics is done in [1] for the hedgehog-like defect and disclinations in the nematic phase of a liquid crystal. The effective metric felt by light in the presence of a hedgehog-like defect is represented by the line element

$$ds^2 = dr^2 + b^2 r^2 \left(d\theta^2 + \sin^2\theta d\phi^2\right), \tag{5}$$

where $b = \frac{n_e}{n_o}$, and in the presence of a $\{\kappa = 1, \gamma = 0\}$ disclination,

$$ds^2 = d\rho^2 + b^2 \rho^2 d\phi^2 + dz^2 \tag{6}$$

Equation (5) is identical to the spatial part of the global monopole's line element, as well eq. (6) is identical to the spatial part of the cosmic string's line element. With this result in hands, the calculation of the light diffraction in the presence of one of each liquid crystal defects will be developed in the next two sections.

### *2.1 The time coordinate: a digression*

A comment must be done. The effective metrics used in this article are riemannian ones, with three spatial coordinates. However, considering the time coordinate, the real world is represented by a four-dimensional pseudo-riemannian metric [12]. To satisfy this, we use the fact that to apply Fermat's principle in a three-dimensional metric, with only spatial coordinates, is equivalent to calculate null geodesics in a static four-dimensional pseudo-riemannian metric [18]. As we are studying static defects, all the riemannian effective metrics treated here in this article can be thought as pseudo-riemannian ones. Therefore, the line elements (5) and (6) must be modified to

$$ds^2 = -dt^2 + dr^2 + b^2 r^2 \left(d\theta^2 + \sin^2\theta d\phi^2\right), \tag{7}$$

$$ds^2 = -dt^2 + d\rho^2 + b^2 \rho^2 d\phi^2 + dz^2. \tag{8}$$

**3 Hedgehog Defect**

In this section, we will apply the method of partial waves [19] to calculate the light diffraction inside a liquid crystal in the nematic phase, being this altered by the presence of a hedgehog-like defect. For simplicity, we will consider the light as a scalar wave, allowing the use of d'Alembert's wave equation:

$$\nabla_\mu \nabla^\mu \Phi \equiv \frac{1}{\sqrt{g}} \partial_\mu \left( \sqrt{-g} g^{\mu\nu} \partial_\nu \Phi \right) = 0, \tag{9}$$

where $g_{\mu\nu}$ are the elements of the metric related to the line element (7) and $g$ is the determinant of $g_{\mu\nu}$.

Considering a plane wave light propagating in the z direction with a given frequency ω,

$$\Phi(t, r, \theta, \phi) = e^{-i\omega t} \psi(r, \theta, \phi), \tag{10}$$

and using (10) in (9), we obtain

$$\left( \Delta - \omega^2 \right) \psi = 0, \tag{11}$$

being Δ written as

$$\Delta = \frac{\partial_r \left( r^2 \partial_r \right)}{r^2} + \frac{\overline{L}}{b^2 r^2},$$

and the operator $\overline{L}$ is the angular part of the Laplacian in flat space in spherical coordinates. At this point, we can use separation of variables in spherical harmonics over $\psi(r, \theta, \phi)$. However, the spherical symmetry implies no dependence on the φ coordinate (*m*=0 for spherical harmonics). Thus, the spatial part for light wave equation is

$$\psi(r, \theta) = \sum_{l=0}^{\infty} a_l R_l(r) P_l(\cos\theta), \tag{12}$$

where $P_l(\cos\theta)$ is the Legendre polynomial of order $l$. As the main quantity obtained from the method of partial waves is the phase shift, the calculation of the term $a_l$ in the previous equation will be ignored. Thus, when substituting (12) in (11), we will have the following radial equation:

$$R_l''(r)+2\frac{R_l'(r)}{r}+\left[\omega^2-\frac{l(l+1)}{b^2r^2}\right]R_l(r)=0,$$

Where $R_l'(r)\equiv\frac{\partial R_l(r)}{\partial r}$. After the change of variables $R_l(r)\equiv\frac{G_l(r)}{\sqrt{r}}$ in the previous equation, we find:

$$r^2G_l''(r)+rG_l'(r)+\left\{\omega^2r^2-\left[\frac{l(l+1)}{b^2}+\frac{1}{4}\right]\right\}G_l(r)=0.$$

The solutions of this equation can be described by first kind Bessel functions [20]

$$G_l(r)=J_{n(l)}(\omega r),$$

where the phase of this function is

$$n(l)=\frac{1}{b}\left[\left(l+\frac{1}{2}\right)^2-\frac{1-b^2}{4}\right]^{\frac{1}{2}}.$$

We observe that when $b=n_e/n_o=1$ we get $n(l)=l+\frac{1}{2}$, representing a wave propagating in flat space, *i.e.*, in the absence of the defect. Comparing the phases of the wave functions in the presence of $\left(\frac{\pi n(l)}{2}\right)$ and in the absence of the defect $\left(\frac{\pi}{2}\left(l+\frac{1}{2}\right)\right)$, we find the following phase shift

$$\delta_l(b) = \frac{\pi}{2}\left(l + \frac{1}{2} - n(l)\right)$$
$$= \frac{\pi}{2}\left(l + \frac{1}{2} - \frac{1}{b}\left[\left(l + \frac{1}{2}\right)^2 - \frac{1-b^2}{4}\right]^{\frac{1}{2}}\right). \qquad (13)$$

In the partial waves method, an important function is $f(\theta)$, called the *scattering amplitude*. Its expression is given by [19]

$$f(\theta) = \frac{1}{2i\omega}\sum_l^{\infty}(2l+1)\left(e^{2i\delta_l} - 1\right)P_l(\cos\theta). \qquad (14)$$

For a given numerical value $b$ at (13) and substituting into (14), the latter equation allows us to calculate numerically the differential scattering cross section $\sigma(\theta)$ (for a numerical example, Fig. 1), *i. e.*, the angular distribution of scattered light, by the following expression [19]

$$\sigma(\theta) = \sigma(\theta,\phi) \equiv |f(\theta)|^2, \qquad (15)$$

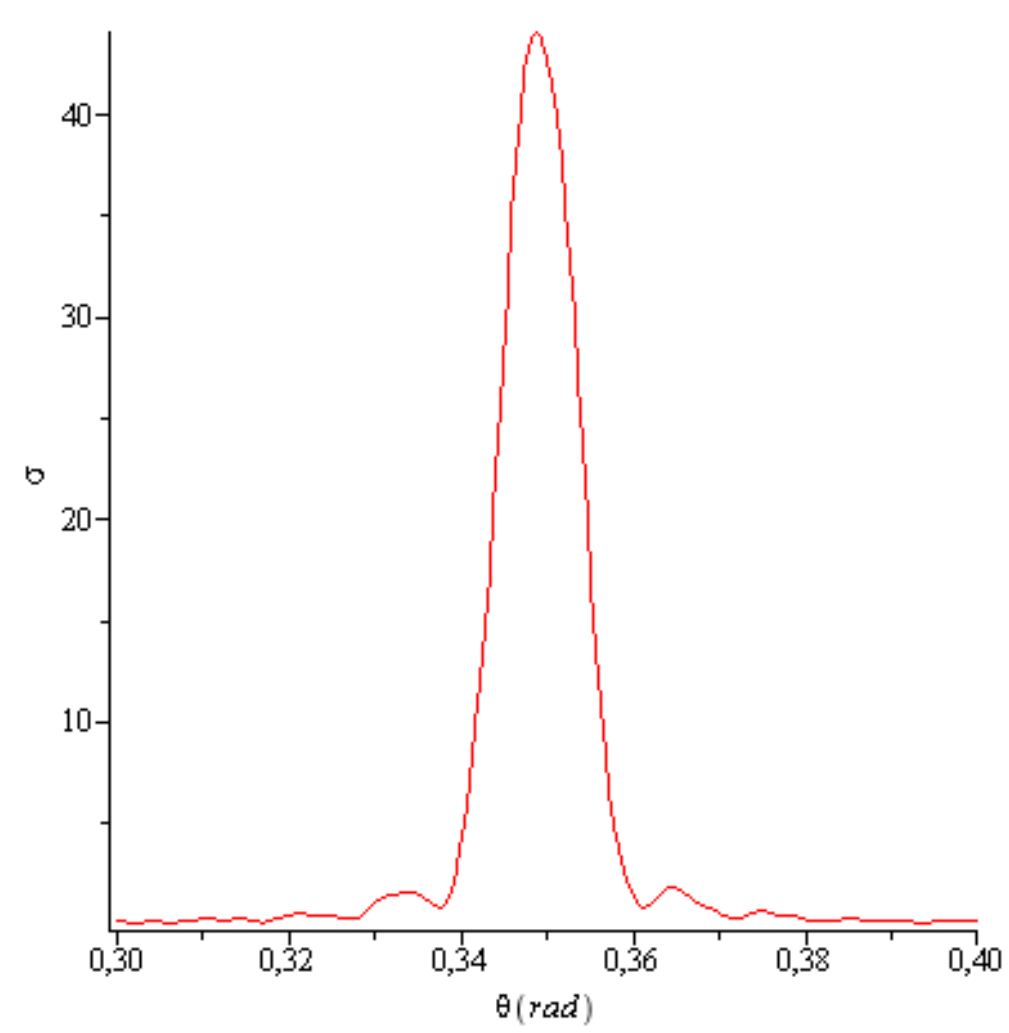

Figure 1. Differential scattering cross section, eq. (15), for a hedgehog with $b$=0.9 using eq. (14) truncated at $l$=600.

where the azimuthal symmetry of the scattering was used. This symmetry generates an annular diffraction pattern, with its maximum located at the angle indicated by Fig. 1. This diffraction ring is the main characteristic of light scattering in the proximity of a global monopole.

Using the method presented in [21], on the $b^2 \approx 1$ regime, we can develop an analytical expression for the angle that locates the diffraction ring and for the total scattering cross section $\sigma_{tot}$ given by the optical theorem [19],

$$\sigma_{tot} = \frac{4\pi}{k}\,\mathrm{Im}\left[f(0)\right] \tag{16}$$

where we made the speed of light $c = 1$ and $k = \omega$.

Initially, making $\zeta \equiv l + \frac{1}{2}$ and $a^2 \equiv \frac{1-b^2}{4}$ in (13), we have

$$\delta_l(b) = \frac{\pi}{2}\left(\zeta - \frac{\zeta}{b}\sqrt{1-\frac{a^2}{\zeta^2}}\right) \tag{17}$$

We expand eq. (17) about and rewrite it in terms of $a^2$, resulting in

$$\delta_l(b) \approx \frac{\pi}{2}\left[\left(1-\frac{1}{b}\right)\zeta + \frac{a^2}{2b\zeta} + O(a^4)\right] \tag{18}$$

As $a^2$ is small, the first two terms of the previous expansion are enough for approximate calculations. We can apply (18) in (14) and find the first and the second term of the expansion $f(\theta) = f^{(0)}(\theta) + f^{(1)}(\theta) + \ldots$ for the scattering amplitude. These two terms can be written through the function

$$h(\theta,\alpha) = \sum_{l=0}^{\infty} e^{i\pi\alpha\zeta(l)} P_l(\cos\theta) = \frac{1}{\sqrt{2(\cos\pi\alpha - \cos\theta)}},$$

related with the generating function for the Legendre polynomials [20], where $\alpha \equiv 1 - \frac{1}{b}$.

The first term of the expansion of the scattering amplitude is proportional to the derivative of $h(\theta,\alpha)$ in relation to α:

$$f^{(0)}(\theta) = \frac{1}{2\sqrt{2}k} \frac{\sin \pi\alpha}{\sqrt{2(\cos\pi\alpha - \cos\theta)}}. \tag{19}$$

Substituting this result in the eq. (16), we find the first term of the expansion $\sigma_{tot} = \sigma^{(0)} + \sigma^{(1)} + \ldots$ :

$$\sigma^{(0)} = \frac{\pi}{k^2} \frac{\cos\frac{\pi\alpha}{2}}{\sin^2 \frac{\pi\alpha}{2}}. \tag{20}$$

The second term of the scattering amplitude, after considering the order $O(a^2)$ in (18), is proportional to $h(\theta,\alpha)$:

$$f^{(1)}(\theta) = \frac{\pi\alpha^2}{2bk} \frac{1}{\sqrt{2(\cos\pi\alpha - \cos\theta)}}. \tag{21}$$

And, again, applying this result in the optical theorem, we find:

$$\sigma^{(1)} = \frac{\pi\alpha^2}{bk^2} \frac{1}{\sin\frac{\pi\alpha}{2}}. \tag{22}$$

As it can be seen, equations (19) and (21) have singular behaviors for the angle $\theta_0 = \pi\alpha$ . Applying this result in (15), we find an annular diffraction pattern located in $\theta_0 = \pi\left(1 - \frac{1}{b}\right)$, valid only for $b^2 \approx 1$. However, when comparing this localization with the one gotten by means of numerical calculation developed by a direct application of (13) in (14), we see that both predict the same position to the diffraction ring for a given *b*. This means that the terms $f^{(0)}(\theta)$ and $f^{(1)}(\theta)$ are indeed predominant in the calculation of the accurate value of $f(\theta)$.

## 4 Disclination

We consider a plane wave light propagating perpendicular for the $\{\kappa = 1, \gamma = 0\}$-disclination line. Due to the problem presents a cyllindrical symmetry, we describe the influence of the disclination as a potential $V(\mathbf{r}) = V(\rho)$. Thus it becomes natural, for the study of the light diffraction, the application of the cylindrical partial waves method. This method will be developed similarly to [21] for the spherical case. The wave problem in this section is related to a defect that is prolonged infinitely in the direction $\hat{z}$, allowing only two space variables $(\rho, \phi)$. Consequently, the wave equation will be given by

$$\left(\Delta - k^2\right)\frac{e^{ik\rho}}{\rho^{1/2}} = 0, \tag{23}$$

where the factor $\rho^{-1/2}$ guarantees that the total flow of energy passing through a circle of radius $\rho$ is independent of ϱ for great values. Like this, the wave function that represents the scattering state, $v_k^{(diff)}(\mathbf{r})$, will be the composition of the plane wave, for instance, spreading along $\hat{x}$, $e^{ikx}$, and a scattered wave (being ignored the problem of the normalization). In this way, the behavior of $v_k^{(diff)}(\mathbf{r})$ for great values of $\rho$ will be

$$v_k^{(diff)}(\vec{r}) \approx e^{ikx} + f_k(\phi)\frac{e^{ik\rho}}{\rho^{1/2}}. \tag{24}$$

In order to apply the method of partial waves, we need three elements. First, the solution of equation (23) in the absence of potential, *i.e.*, the free cylindrical wave $\psi_{k,l}^{(0)}(\rho, \phi)$,

$$\psi_{k,l}^{(0)}(\rho,\phi) \approx J_l(k\rho)e^{il\phi}$$
$$\lim_{\rho\to\infty} \psi_{k,l}^{(0)}(\rho,\phi) \approx \frac{1}{\sqrt{k\rho}} \cos\left(k\rho - \frac{\pi}{2}\left(l - \frac{1}{2}\right)\right)e^{il\phi}. \tag{25}$$

Second, the solution of the equation (23) in the presence of a symmetrical potential $V(\mathbf{r})=V(\rho)$, *i.e.*, the cylindrical partial waves $\psi_{k,l}(\rho,\phi)$,

$$\psi_{k,l}(\rho,\phi) \approx J_{\nu_l}(k\rho)e^{il\phi}$$
$$\lim_{\rho\to\infty} \psi_{k,l}(\rho,\phi) \approx \frac{1}{\sqrt{k\rho}} \cos\left(k\rho - \frac{\pi}{2}\left(\nu_l - \frac{1}{2}\right)\right)e^{il\phi} \tag{26}$$
$$= \frac{1}{\sqrt{k\rho}} \cos\left(k\rho - \frac{\pi}{2}\left(\nu_l - \frac{1}{2}\right) + \delta_l\right)e^{il\phi},$$

being $\delta_l$ the phase shift regarding the free wave (25). And third, the plane wave expanded in free cylindrical waves, the Jacobi–Anger's expansion [22],

$$e^{ikx} = e^{ik\rho\cos\phi} = \sum_{l=-\infty}^{\infty} i^l J_l(k\rho)e^{il\phi} = \sum_{l=-\infty}^{\infty} i^l \varepsilon_l J_l(k\rho)e^{il\phi}, \tag{27}$$

where $\varepsilon_0 \equiv 1$ and $\varepsilon_l \equiv \left(1 + e^{-2il\phi}\right)$ $(l \geq 1)$.

With these three elements, it is possible to expand $v_k^{(diff)}(\mathbf{r})$ in a series of cylindrical partial waves in such a way that the coefficients of (27) will be used so that $v_k^{(diff)}(\mathbf{r})$ reproduces a plane wave when $\rho \to \infty$:

$$v_k^{(diff)}(\vec{r}) = \sum_{l=0}^{\infty} c_l \psi_{k,l}(\rho,\phi) = \sum_{l=0}^{\infty} i^l \varepsilon_l \psi_{k,l}(\rho,\phi). \tag{28}$$

For $\rho \to \infty$ and using the exponential form of the function cosine, the substitution (26) in (28) results in

$$v_k^{(diff)}(\vec{r}) = \sum_{l=0}^{\infty} i^l \varepsilon_l e^{il\phi} \frac{e^{i\left(k\rho - \frac{\pi}{2}\left(l - \frac{1}{2}\right)\right)} e^{i\delta_l} + e^{-i\left(k\rho - \frac{\pi}{2}\left(l - \frac{1}{2}\right)\right)} e^{-i\delta_l}}{2\sqrt{k\rho}}.$$

Multiplying the right-hand side of the last equation by the factor without physical meaning $e^{i\delta_l}$ and using $e^{2i\delta_l} = 1 + 2ie^{i\delta_l} \sin\delta_l$, one attains

$$v_k^{(diff)}(\vec{r}) = \sum_{l=0}^{\infty} i^l \varepsilon_l e^{il\phi} \left( \frac{e^{i\left(k\rho - \frac{\pi}{2}\left(l - \frac{1}{2}\right)\right)} e^{i\delta_l} + e^{-i\left(k\rho - \frac{\pi}{2}\left(l - \frac{1}{2}\right)\right)} e^{-i\delta_l}}{2\sqrt{k\rho}} + \frac{e^{i\left(k\rho - \frac{\pi}{2}\left(l - \frac{1}{2}\right)\right)} 2ie^{i\delta_l} \sin\delta_l}{2\sqrt{k\rho}} \right).$$

Recognizing the first term between parentheses as being the asymptotic form of (27) and putting in evidence the term $\frac{e^{ik\rho}}{\rho^{1/2}}$ in the second term, the comparison of this with (24) identifies $f_k(\phi)$ as being

$$f_k(\phi) = \frac{1}{\sqrt{k}} \sum_{l=0}^{\infty} ie^{\frac{-i\pi}{4}} \varepsilon_l e^{il\phi} \sin\delta_l \, e^{i\delta_l}. \tag{29}$$

Applying (29) in (15) we have the differential scattering cross section for cylindrical defects with angular symmetry.

Now we apply the theory developed so far to calculate the scattering amplitude and the differential scattering cross section for the light in the plane $z$ = *const*. scattered by a disclination with effective metric represented by (6). The configuration of molecules in the plane $z$ = *const*. is similar to the case of the hedgehog defect seen from its equatorial section $\left(\theta = \frac{\pi}{2}\right)$, that can be verified comparing the equations (5) and (6). In this manner, our starting point will be again the application of the effective metric for the disclination (6) in the wave equation (9).

Due to the symmetry in the direction $z$, we will just study the propagation along the coordinates $\rho$ and $\phi$ (a different approach of the similar problem of the cosmic string can be seen in [23]). Therefore, using a separation of variables similar to (10), we found the following wave equation in the presence of a disclination

$$\left(\frac{\partial^2}{\partial r^2}+\frac{1}{r}\frac{\partial}{\partial r}+\frac{1}{b^2 r^2}\frac{\partial^2}{\partial \phi^2}+\omega^2\right)\psi(\rho,\phi)=0.$$

where it will be used $\psi(\rho,\phi)=\sum_{l=0}^{\infty} c_l R_l(\rho)e^{il\phi}$. With the latter expansion, the equation to the radial coordinate will be

$$\left(\frac{\partial^2}{\partial r^2}+\frac{1}{r}\frac{\partial}{\partial r}-\frac{\upsilon_l^2}{r^2}+\omega^2\right)R(\rho)=0,$$

being $R(\rho)=J_{\nu_l}(\rho), \nu_l=\frac{l}{b}$. Using this index, one can find the phase shift on the way to equation (26) and, consequently, the scattering amplitude (29) and the differential scattering cross section (15).

The phase shift of the scattered wave by the disclination (6) is found to be $\delta_l=\frac{l\pi}{2}\left(1-\frac{1}{b}\right)$. Fig. 2 shows the differential scattering cross section with the same parameters of the Fig 1.

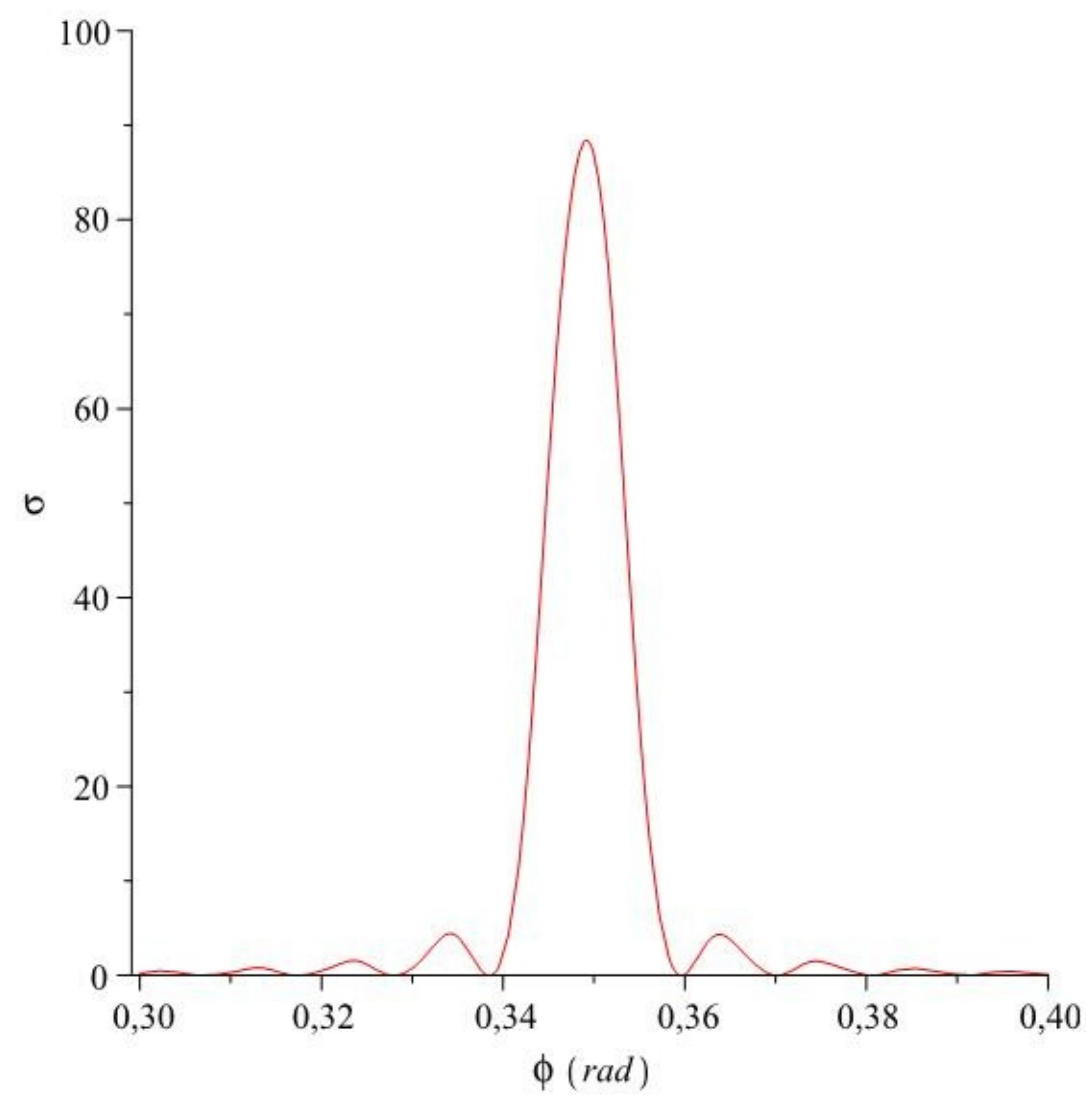

Figure 2. Differential scattering cross section, eq. (15), for a $\{\kappa = 1, \gamma = 0\}$ disclination with $b = 0.9$ using eq. (29) truncated at $l = 600$.

Analogous to the latter, Fig. 2 indicates, for the disclination case, a singular behavior of (15) around $\phi_0 = \pi\alpha$, where $\alpha \equiv \left(1 - \frac{1}{b}\right)$, generating a luminous line representing the diffraction pattern about $\phi_0$. Like this, one could imagine the disclination as a line of punctual hedgehog-like defects, or from the gravitational point of view, a cosmic string as a line of global monopoles.

As observed by Grandjean in 1919 [11, 24], the angle formed by the light ray scattered to the right of the disclination and the light ray scattered to the left of the disclination is $2\pi\left(1 - {n_o}/{n_e}\right)$. This result, which was obtained by a very different method, is exactly the same we obtained in this paper, since $\phi_0$ is merely the angle between the direction of the incoming light beam and the light scattered to one of the sides, because each side is equivalent on account of the azimuthal symmetry. Thus $\phi_0$ is half of the result obtained by Grandjean.

## 5 Temperature Dependence

It is well known the temperature dependence that the refraction indexes of a liquid crystal molecule have [25, 26]. We will use this fact to calculate the influence of the temperature in the position of the diffraction pattern for the previously studied defects.

As exposed in [26, 27], the refractive indexes for ordinary and extraordinary rays traveling in a liquid crystal molecule can be treated by a four-parameter model. Thus, the refraction indexes are expressed by

$$n_e(T) \approx A - BT + \frac{2(\Delta n)_0}{3}\left(1 - \frac{T}{T_c}\right)^{\beta} \tag{30}$$

$$n_o(T) \approx A - BT - \frac{(\Delta n)_0}{3}\left(1 - \frac{T}{T_c}\right)^{\beta}, \tag{31}$$

where $A$ is a dimensionless constant, $B$ is a constant with $(K)^{-1}$ unit, $(\Delta n)_0 \equiv (n_e - n_o)_0$ is the birefringence at $T = 0$ K, β is a material constant and $T_c$ is the transition temperature between the isotropic and nematic phase. Using this model for 5CB (4-cyano-4-n-pentylbiphenyl), that has $T_c = 306.6$ K, and a light beam with wavelength $\lambda = 589$nm, one obtains $A = 1.7674$ and $\beta = 0.1889$ [24]. Applying (30) and (31) in (5) and (6), one obtains a new expression for $b$ and, consequently, the temperature dependence of the diffraction pattern of a hedgehog-like and disclination defect by the relation $\phi_0 = \pi\left(1 - \frac{1}{b}\right)$

.

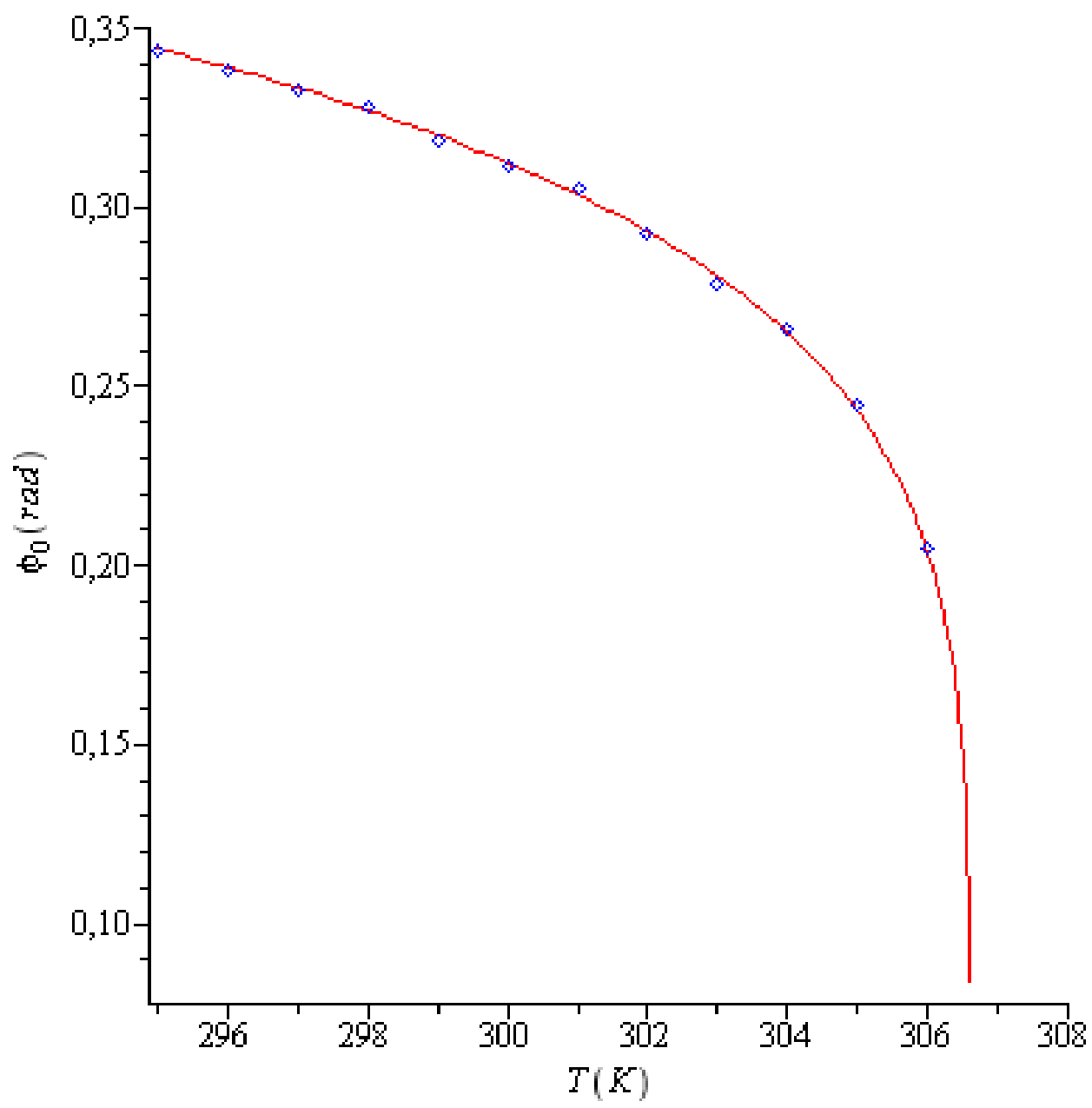

Figure 3. Temperature dependence for the position of diffraction pattern, $\phi_0 = \pi\left(1 - \frac{1}{b}\right)$, for 5CB. The dots are numerical results.

Fig. 3 shows the behavior of the diffraction pattern position for 5CB in its nematic phase in the open interval $T = [295; 306.6]$ K [11] regarding the two previously studied defects. One observes the low sensibility on the position at the beginning of the temperature range ($\approx 295$K) and a high sensibility at the end of the temperature range ($\approx 306$K), near to the nematic-isotropic phase transition.

It's worth to mention that the line in Fig. 3 was obtained by $\phi_0 = \pi\left(1 - \frac{1}{b}\right)$, that came from the approximatation $b^2 \approx 1$, eq. (18), while the dots came from the numerical calculation.

## 6 Conclusion

The relationship between the Fermat's principle and its description by an effective metric allows us to acquire information about the optical properties of the liquid-crystalline material, calculating the position of the diffraction pattern resultant from the light scattering over a hedgehog-like or disclination defect in the nematic phase, because $\theta_0 = \pi\left(1 - \frac{1}{b}\right)$, where $b = \frac{n_e}{n_o}$. With these two relations, it is possible to measure the temperature by the diffraction pattern localization of the light scattered by hedgehog-like or disclination defects, mainly near the high temperature end of nematic phase. It is worthy to note that the method developed here is simpler than others and it is consistent with some experimental results that can be found in the literature (for the hedgehog, a theoretical and experimental study can be seen in [13] and [28], respectively).

This work was partially supported by INCT-CX and CAPES (PROCAD).